\documentclass{sigchi}

\usepackage{balance}  
\usepackage{graphics} 
\usepackage{times}    
\usepackage{url}      
\usepackage{censor}

\usepackage{tabularx}
\usepackage{array}
\usepackage{booktabs} 
\usepackage{multirow}
\usepackage{graphicx}
\usepackage{colortbl}
\usepackage{color}
\newcommand{\up}[1]{\ensuremath{^{\textrm{\tiny#1}}}}

\makeatletter
\def\url@leostyle{%
  \@ifundefined{selectfont}{\def\UrlFont{\sf}}{\def\UrlFont{\small\bf\ttfamily}}}
\makeatother
\def\pprw{8.5in}
\def\pprh{11in}

\usepackage[pdftex]{hyperref}
\hypersetup{
pdftitle={Friend Inspector: A Serious Game to Enhance Privacy Awareness in Social Networks},
pdfauthor={LaTeX},
pdfkeywords={SIGCHI, proceedings, archival format},
bookmarksnumbered,
pdfstartview={FitH},
colorlinks,
citecolor=black,
filecolor=black,
linkcolor=black,
urlcolor=black,
breaklinks=true,
}

\begin{document}

\title{Friend Inspector: A Serious Game to Enhance Privacy Awareness in Social Networks}

\numberofauthors{1}
\author{
  \alignauthor Alexandra Cetto, Michael Netter, G\"{u}nther Pernul, Christian Richthammer, Moritz Riesner, Christian Roth, Johannes S\"{a}nger \\
    \affaddr{Department of Information Systems}\\
    \affaddr{University of Regensburg}\\
    \email{firstname.lastname@ur.de}\\
}



\maketitle

\begin{abstract}
Currently, many users of Social Network Sites are insufficiently aware of who can see their shared personal items. Nonetheless, most approaches focus on enhancing privacy in Social Networks through improved privacy settings, neglecting the fact that privacy awareness is a prerequisite for privacy control. Social Network users first need to know about privacy issues before being able to make adjustments. In this paper, we introduce Friend Inspector, a serious game that allows its users to playfully increase their privacy awareness on Facebook. Since its launch, Friend Inspector has attracted a significant number of visitors, emphasising the need for better tools to understand privacy settings on Social Networks.

\end{abstract}






\section{Introduction}

Over the last decade, Social Network Sites (SNSs) have gained importance as a medium for social interaction, allowing people to stay in touch with existing contacts and to create new relationships. Hereunto, SNSs ease social interaction by offering a centralised point to communicate with contacts from different social spheres (e.g. family members, close friends, and colleagues).

Despite these positive social outcomes, the rise of SNSs has been accompanied by privacy concerns. Besides the broadly discussed SNS service providers' handling of personal data, privacy is also threatened by a SNS user's contacts (often referred to as "friends") \cite{Ziegele2011}. On general-purpose SNSs such as Facebook, "unimaginably complex social relations collapse to the infinitely thin plane of a single profile" \cite{Peterson2010}. As a result, it is difficult for a SNS user to simultaneously meet the expectations and respect varying social norms of conflicting social spheres \cite{Binder2009}. This might put the user at risk of offending one (or more) of these social spheres, ultimately leading to social exclusion. A SNS user, for instance, may struggle with targeted sharing sensitive family-related pictures with close friends and family members while hiding these pictures from his colleagues who have also access to his SNS profile. Generally speaking, privacy is threatened if shared personal items are visible to contacts for whom they are not intended. 

However, these privacy issues are not primarily due to a lack of appropriate privacy settings, as popular SNSs offer a wide range of fine-grained controls to adjust the visibility of shared items \cite{Riesner2013}. Instead, it has been shown that an item's visibility is often only defined once when it is shared and subsequently left unchanged \cite{Strater2008}. Over time and due to the large number of shared items and contacts, users become unaware of who has access to which shared items \cite{Netter2013}. Awareness of inaccurate privacy settings, however, is a prerequisite for being able to make necessary changes. Put differently, users first need to know of misconfigured privacy settings before being able to make adjustments. 

Especially for young people, a careless attitude towards SNS privacy puts their future prospects (such as when applying for a job) at risk and may lead to social exclusion. On the one hand, the age group of people between 13 and 25 is most active on SNSs. On the other hand, the "cyber personae they spawned in adolescent efforts to explore identity have taken on permanent lives in the multiple archives of the
digital world." \cite{Rosenblum2007} Hereunto, early and playful education of privacy risks on SNSs can contribute to responsible usage and empower those people to harness the strengths of SNSs. In this paper, we adopt the concept of serious games in order to strengthen privacy awareness on SNSs. It has been widely accepted that games can provide an engaging and motivational environment for learning \cite{Kiili2005}. In \cite{Freitas2011}, the efficacy of game-based approaches for behavioural change has been demonstrated. Besides, it has been shown that serious games have the potential to increase awareness of important societal issues \cite{Rebolledo-Mendez2009}. Our resulting serious game, termed \textit{Friend Inspector}, is a browser-based application that allows Facebook users to playfully check their knowledge of who can see their shared personal items and provides personalised recommendations on how to improve privacy settings.

The remainder of this paper is structured as follows. After examining related work in the following section, an in-depth discussion of the concepts of privacy and serious games as the two foundations of Friend Inspector is provided. Based thereupon, the conceptual design of Friend Inspector is presented. Finally, we discuss implementation details and conclude the paper.
\section{Related Work}
\label{relwork}

Raising privacy awareness on SNSs has been the subject of both practical and theoretical approaches. Practical approaches such as Profile Watch\protect{\footnote{http://www.profilewatch.org/}}, Privacy Check\protect{\footnote{http://www.rabidgremlin.com/fbprivacy/}}, and Privacy Scanner\protect{\footnote{http://www.reclaimprivacy.org/}} analyse privacy settings and publicly shared items of a Facebook profile. Subsequently, results are summarised and recommendations that offer guidance on how to improve privacy on Facebook are provided. Unfortunately, two of the three approaches were not fully functional as of November 2\up{nd}, 2013. Friend Inspector differs from these approaches as these sites only analyse publicly available items instead of all shared items. From an educational perspective, due to their informatory-only approach, these sites are of limited value for sustained learning compared to Friend Inspector, which allows users to actively test their knowledge of their SNS privacy settings in a playful manner.

Aside from tools to check privacy settings, game-based approaches such as Realistic Facebook Security Simulator\protect{\footnote{http://toys.usvsth3m.com/realistic-facebook-privacy-simulator/}} and Privacy Game\protect{\footnote{http://www2.open.ac.uk/openlearn/privacy/game/}} exist. Realistic Facebook Security Simulator presents a set of privacy-related questions, asking the user to specify his prefered visibility for typical information shared on SNSs within a limited amount of time. At the end of each round, the answers are evaluated. The user passes to the next round if the answers were correct, otherwise the game is over. Unlike Friend Inspector, this game-based approach does not employ items from the user's Facebook profile but solely operates on a static set of questions, making it difficult for users to relate the results to actual privacy issues on their profiles. Besides, Privacy Game is a general-purpose game for privacy education. The game operates on a set of predefined information pieces, requiring each player to make decisions whether to reveal particular information such as by trading it for gifts on a shopping site. Privacy Game differs from Friend Inspector in several ways. On the one hand, it is not specifically designed to raise privacy awareness in SNSs. On the other hand, the game lacks personalisation as it uses fictitious information pieces rather than using the player's actual personal items from his SNS profile.

From the viewpoint of academic research, raising security awareness has been the focus of few game-based approaches such as CyberCIEGE \cite{Irvine2005} and Control-Alt-Hack \cite{Denning2013}. However, their objective is to playfully learn about security in general. Unlike Friend Inspector they do neither specifically focus on the awareness aspect of privacy nor on SNSs in particular. Besides these two games, few visualisation-based approaches to increase privacy awareness on SNSs exist that provide a readily understandable presentation of privacy settings \cite{Lipford2008, Lipford2010, Anwar2010, Anwar2011}. While offering different views on the visibility implications of privacy settings, these approaches are only of informing nature and lack any game-based elements.

\section{Background}
\label{background}
In this section, we explicate the concepts of privacy in SNSs and serious games as the two foundations of Friend Inspector in order to arrive at a common understanding. Both foundations significantly influence the learning objectives as well as the design of the proposed game.

\subsection{Privacy in SNSs}

In order to develop a game-based approach to improve privacy awareness on SNSs, first a common understanding of both terms (privacy and awareness) is needed. 

Literature offers a variety of privacy conceptualisations such as \textit{control over personal information} \cite{Solove2002}, \textit{confidentiality or secrecy of personal information} \cite{CommitteeonPrivacyintheInformationAgeNationalResearchCouncil2007}, or \textit{freedom to construct one's identity} \cite{Agre1998}. For this work, we build upon Nissenbaum's view on privacy as \textit{contextual integrity} \cite{Nissenbaum2010}, which is commonly used to understand privacy issues in an environment of voluntary information disclosure such as SNSs \cite{Hull2011}. Privacy in the sense of contextual integrity is about respecting the social norms (established by culture, history, and conventions) in a given situation (context). Based on this definition, sharing personal information per se is not a privacy issue \cite{Leenes2010}. Privacy is only threatened if this information is shared outside the context in which it was initially shared. Applying contextual integrity to SNSs, sharing family-related pictures with one's parents, for instance, preserves contextual integrity and does not violate privacy. In contrast, privacy is violated if such pictures leave the intended context and become available to one's colleagues or employer.

\begin{figure}
\centering
\includegraphics[width=1.0\columnwidth]{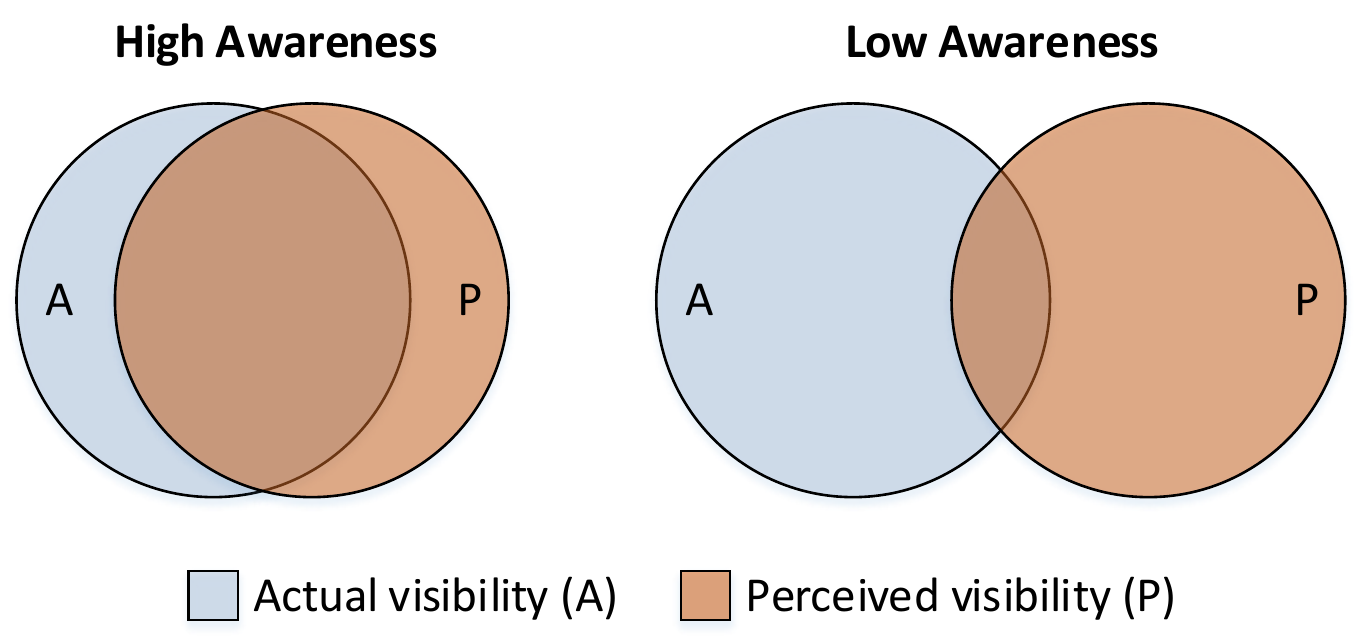}
  \caption{Privacy Awareness (Venn diagram)}
\label{fig:privacyawareness}
\end{figure}

Privacy awareness on SNSs can be defined as an individual's knowledge of who can access which shared personal information. In more detail, privacy awareness is the degree to which \textit{actual} and \textit{perceived} visibility of shared items match \cite{Netter2013}. Figure \ref{fig:privacyawareness} illustrates privacy awareness using set theory. Two sets can be defined: perceived visibility settings $P$ and actual visibility settings $A$. A user is highly privacy aware if the sets $A$ and $P$ largely intersect, i.e. the perception of who can access which items largely corresponds to what is defined on the SNS. Likewise, a minor or no intersection of $A$ and $P$ implies a low privacy awareness.

Currently, several factors have a negative impact on privacy awareness on SNSs. Firstly, users typically share a large number of items on SNSs with a large number of contacts\footnote{For instance, the average Facebook user has 190 contacts \cite{Ugander2011}.}. As a consequence, it becomes increasingly difficult after a while to remember which contacts can see which personal items. Secondly, it is highly context-dependent whether or not an item is considered private. A user's shared picture of him being drunk at a party, for instance, may not be considered highly sensitive from a close friend's perspective but is highly private to the user with regard to potential future employers and may lead to social exclusion. Lastly, current SNSs are optimised for information sharing rather than for gaining privacy awareness \cite{Lipford2008}. Hence, existing means to review visibility settings are tedious to use as they often require a user to manually check each shared item.

As a result of the previous discussion, the following two privacy-related objectives can be derived for the design of a privacy awareness game:
\begin{itemize}
	\item \textbf{Preselection of Sensitive Items:} Reduce complexity by focusing only on few privacy-relevant items. As privacy is context-dependent, players themselves must specify the sensitivity of their items.
	\item \textbf{Comparison of Actual and Perceived Visibility:} For the selected items, provide simple means to compare a player's perceived visibility with the actual visibility.
\end{itemize}


\subsection{Serious Games}

It has been widely accepted that games can provide an engaging and motivational environment for learning \cite{Kiili2005}. While entertainment can be seen as the main motivation of traditional games, serious games that combine both computer and video games for non-entertainment purposes have become popular in the last decade \cite{Marsh2011}. A precise definition of the notion of serious games is still difficult to formulate due to rapid technological and artistic developments and innovations made in the virtual and gaming environments \cite{Marsh2011}. For this work we use Marsh's definition who tried to fill that gap:

\textit{``Serious games are digital games, simulations, virtual environments and mixed reality/media that provide opportunities to engage in activities through responsive narrative/story, gameplay or encounters to inform, influence, for well-being, and/or experience to convey meaning. The quality or success of serious games is characterised by the degree to which purpose has been fulfilled. Serious games are identified along a continuum from games for purpose at one end, through to experiential environments with minimal or no gaming characteristics for experience at the other end.''} \cite{Marsh2011}


In this work, we focus on digital game-based learning in an experiential environment as one aspect of the serious games continuum. In experiential environments, an inductive learning approach is used. By contrast, traditional instructional design usually includes methods to encourage deductive learning \cite{Amory2003}. Thereby, a concrete concept or solution is presented followed by exercises to practice. Inductive learning, in contrast, is based on discovery. It allows students to "invent" a solution or concept by experimentation which is often considered a more effective approach \cite{Prince2006}.   

\begin{figure}
\centering
\includegraphics[width=3.40in]{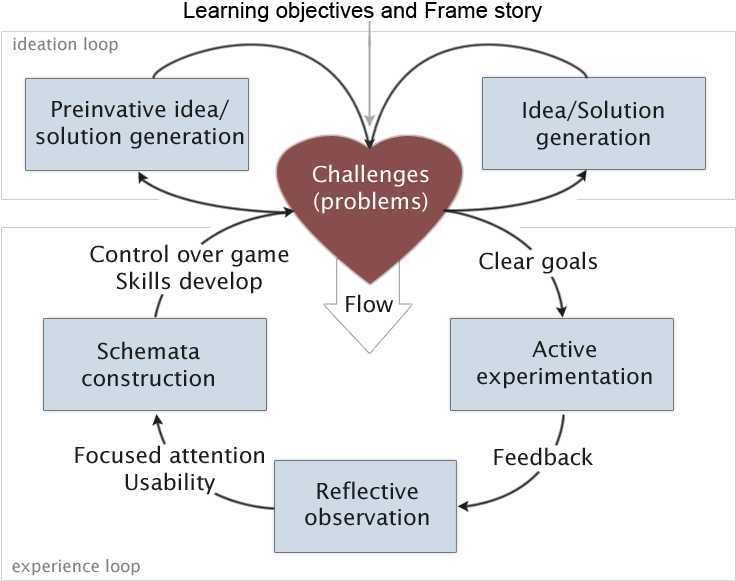}
\caption{Experiential gaming model (based on \protect\cite{Kiili2005})}
\label{model}
\end{figure}

A framework that tries to combine experiential learning theory (inductive), flow theory and game design is the experiential gaming model proposed in \cite{Kiili2005}. The experiential gaming model depicted in Figure \ref{model} on which Friend Inspector is based ``describes learning as a cyclic process through direct experience in the game world'' \cite{Kiili2005}. The starting point of the experiential gaming model are the learning objectives. Based on these, a player is presented with one or several challenges / problems. While solving these problems, the learner runs through a cyclic process that involves both an experience loop where the learner conducts active experimentation, reflective observations, and schemata construction and an ideation loop where ideas or solutions are generated. Although the experiential gaming model works as a tie between educational theory and game design, it does not cover the whole gaming process. Thus, a frame story is needed that integrates the challenges into a larger task or a problem \cite{Kiili2005}.

Based on this, the following two components / concepts have to be elaborated during the design process:

\begin{itemize}
	\item \textbf{Frame story}: Provide a frame story that motivates the learner and integrates the challenges into a meaningful context.
	\item \textbf{Experiential gaming model}: Design the learning phase using the experiential gaming model to achieve the learning objectives.
\end{itemize}


\section{Conceptual Design of Friend Inspector}
\label{concept}

\begin{figure*}
\centering
 \includegraphics[width=\linewidth]{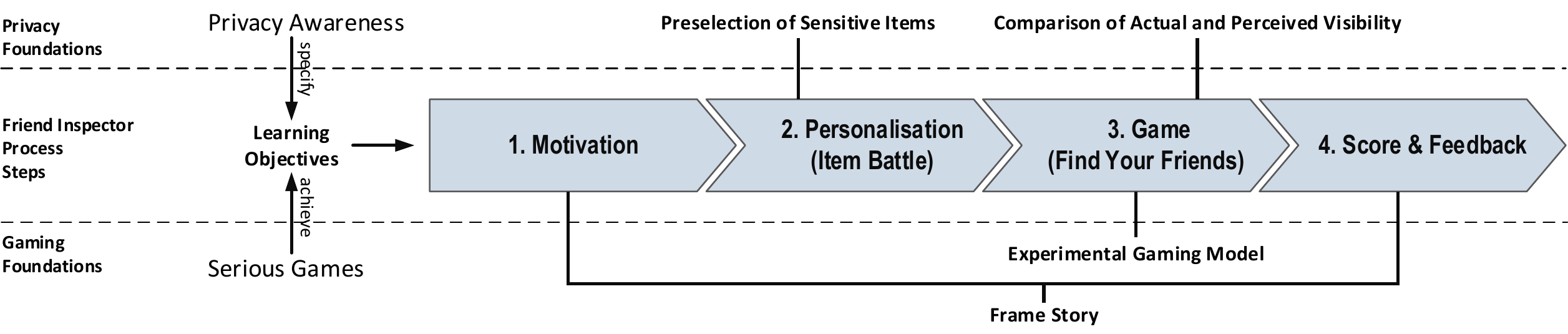}
\caption{Friend Inspector Process Flow}
\label{process}
\end{figure*}

Based on the foundations of privacy awareness and serious games, forming the pillars of our work, in this section we present the design of Friend Inspector. The design has been iteratively improved during several refinement cycles.

\subsection{Frame story and learning objectives}
The general idea of Friend Inspector is to create an environment in which users can playfully discover their privacy settings on Facebook and find out who can see their shared items. Based on this, they can ultimately increase their privacy awareness. To this end, the central element of Friend Inspector is a memory-like game, where players are asked to guess the visibility of a presented item within a limited amount of time. A score is calculated based on the correctness of the answer and the time needed to complete the task. 

In order to provide a meaningful context for this challenge, the frame story is to motivate SNS users to test their privacy awareness about items they shared on their Facebook profile and to reach the best possible score. To increase the game's competitive element, the score can be shared on the user's Facebook wall to challenge other contacts to beat his score. As a user's social value on SNSs is created by sharing interesting things, sharing a Friend Inspector score creates a positive feedback loop in which people in a user's network of contact mutually try to beat each other and share a higher score. 

Based on the previous analysis of privacy awareness in SNSs, the following two learning objectives should be reached:
\begin{itemize}
	\item \textbf{Enhance privacy awareness}: We want users to recognise the properties and consequences of their privacy settings. Thereby, we want to decrease the gap between perceived and actual visibility.
	\item \textbf{Learn about privacy settings}: By giving recommendations, we want to empower users to improve their privacy settings based on their desired preferences.
\end{itemize}

\subsection{Process flow}
To reach the learning objectives, we designed Friend Inspector as a four-step process flow (see Figure \ref{process}). The process flow integrates the concepts of \textit{privacy awareness}, which specifies the learning objectives, and \textit{serious games} in order to achieve these objectives. Each step of the four-step process flow is preceded by a briefing window that provides the learner with basic instructions. The first step (Motivation) and the last step (Score \& Feedback) form the frame story, integrating the challenges into a meaningful context that invites a Facebook user to play Friend Inspector. Step two (Personalisation) is used to reduce the total number of items for the subsequent game step and elicit those items the user considers especially sensitive. Step three (Game) contains the main element of Friend Inspector, building upon the experimental gaming model. During this step, the user has to guess the visibility of a presented item by selecting the respective people from a set of depicted contacts. Step three is repeated five times before the game continues with the final Score \& Feedback step. 

In the following sections, the conceptual design of each step is presented in detail. 

\begin{figure}[!h]
\centering
\includegraphics[width=3.40in]{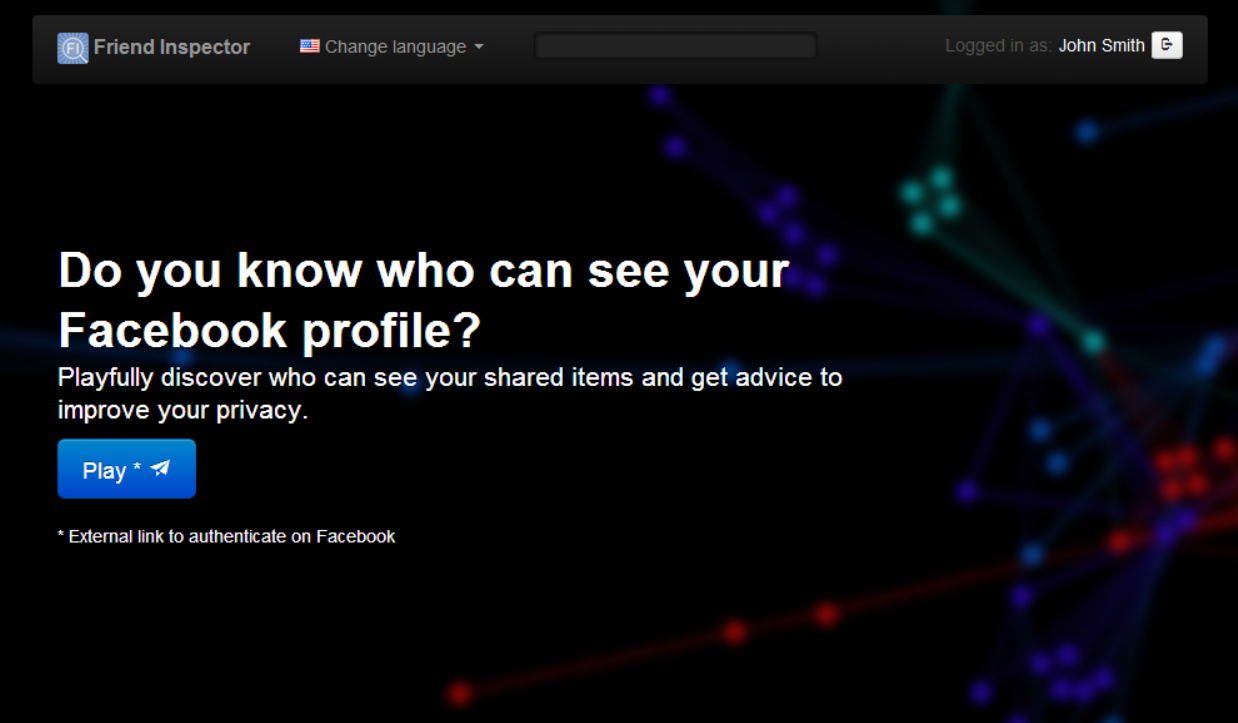}
\caption{Friend Inspector landing page (step 1)}
\label{step1}
\end{figure}

\subsubsection{Motivation step}
Figure \ref{step1} depicts Friend Inspector's initial landing page. At the very beginning of the game, a potential user is teased by the question ``Do you know who can see your Facebook profile?''. As motivation is important in this context, the objective of this phrase is to gain the users' attention and invite them to test their knowledge of the privacy settings of their Facebook profile. The subtopic ``Playfully discover who can see your shared items and get advice to improve your privacy.'' aims to further clarify the objectives and their relevance. It hints at the underlying game-based approach that distinguishes Friend Inspector from purely educational or information giving approaches such as presented in the related work section. The design of the landing page follows the first two components of Keller's ARCS (attention, relevance, confidence, satisfaction) model for motivational design, using a question as attention strategy (inquiry) and the worth for the user as relevance strategy (presented worth) \cite{Keller1987}.

\subsubsection{Personalisation step (Item Battle)}
Step two is concerned with adapting the game to the player's Facebook profile. Personalisation is crucial, as adaption to learners is considered a key aspect for the success of serious games \cite{Romero2012}. In the context of Friend Inspector, personalisation deals with integrating the player's own items and their privacy settings into the game and preselecting sensitive items.

\begin{figure}[!t]
\centering
\includegraphics[width=3.40in]{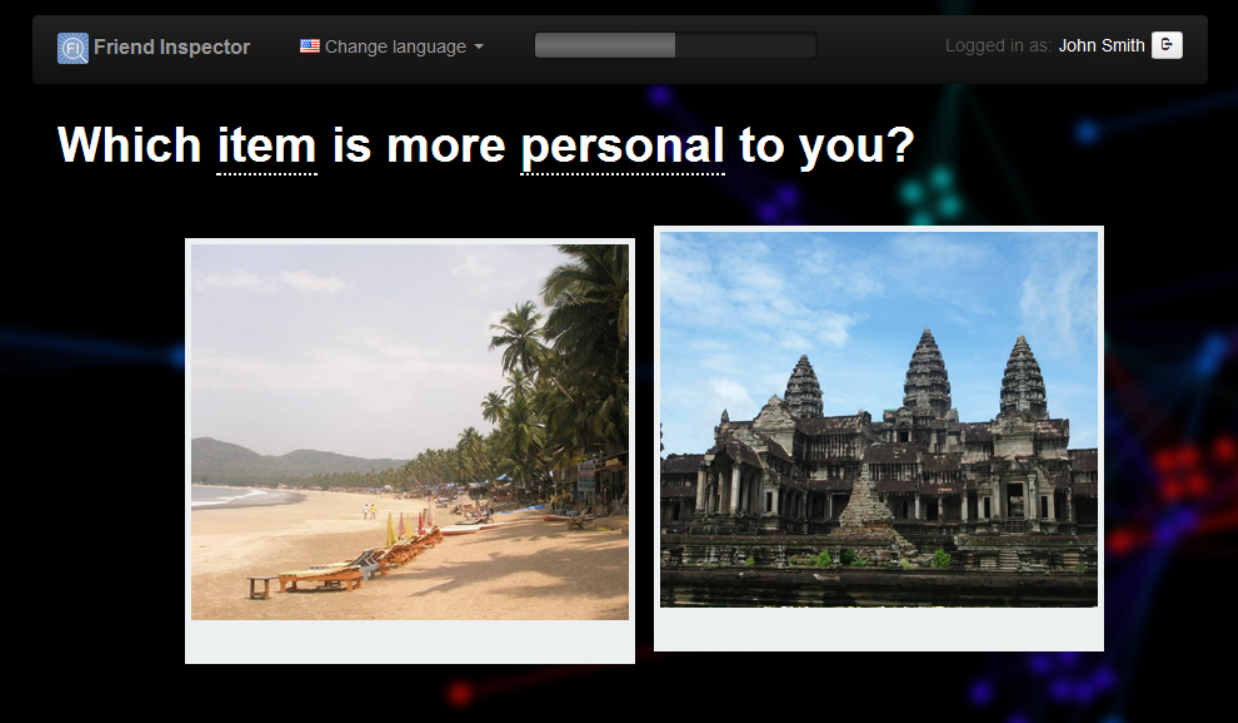}
\caption{Item Battle (step 2)}
\label{step2}
\end{figure} 

To perform the personalisation, Friend Inspector initially retrieves a user's contacts as well as his shared items and the corresponding privacy settings from Facebook. Subsequently, sensitive items (for which the user has strong visibility preferences) need to be determined from the set of all shared items. Friend Inspector offers a playful way to determine the sensitivity of items. To this end, a pair-wise comparison of two displayed items is used, asking the user to select one of the two items which is more personal to him\protect{\footnote{Friend Inspector uses the Elo rating \cite{Elo2008} to rank items.}}. This comparison, termed Item Battle, is executed in ten rounds with varying items and implicitly results in an ordered list of items ranked by sensitivity. Figure \ref{step2} shows Item Battle for two exemplary items. A subset of the most sensitive items is then used during the actual gaming step.

\subsubsection{Game step (Find Your Friends)}
\begin{figure}[!h]
\centering
\includegraphics[width=3.40in]{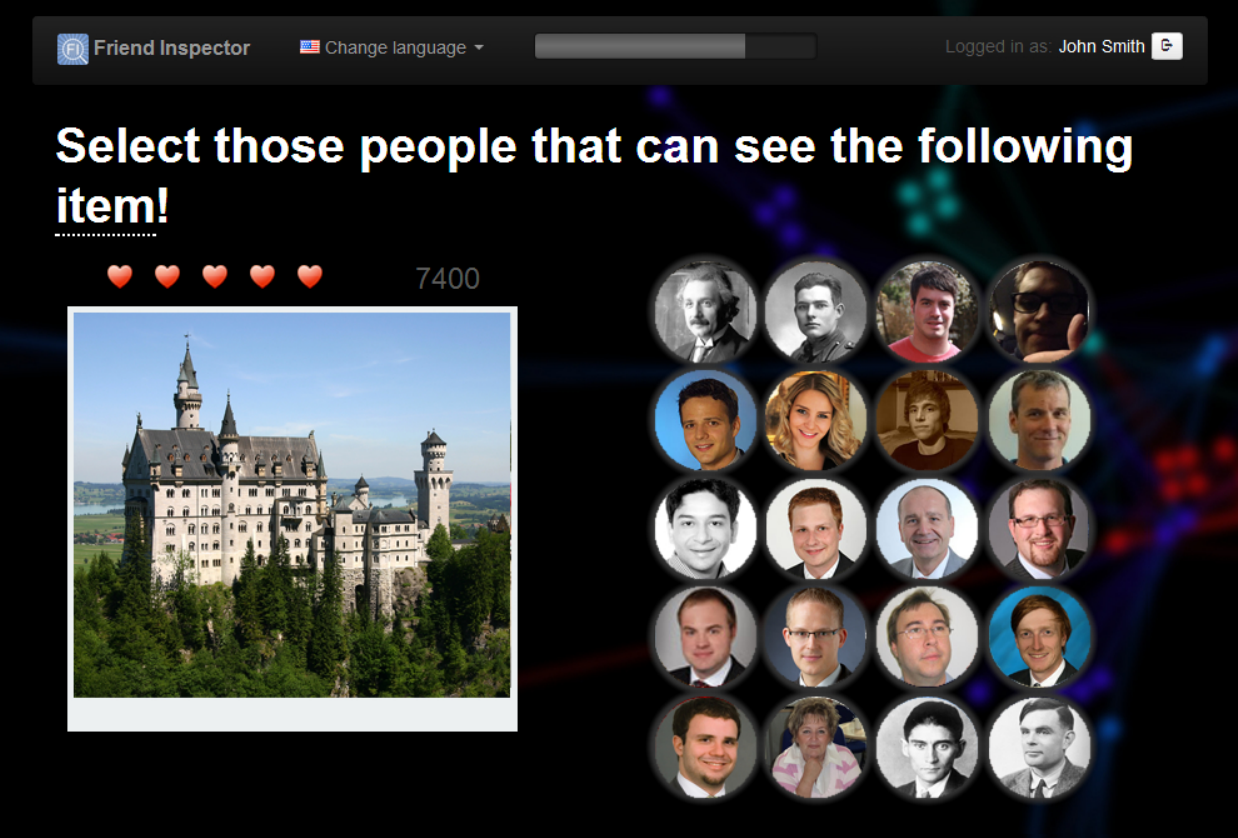}
\caption{Find Your Friends (step 3)}
\label{step3}
\end{figure}

The third step of the Friend Inspector process flow contains the actual game which is termed Find Your Friends. The game consists of five rounds, whereas Figure \ref{step3} depicts the interface of a single round of Find Your Friends. The left area shows one of the user's sensitive items that have been determined in the previous step. The right area contains a set of 20 profile pictures that consist of the user's contacts as well as randomly selected strangers\footnote{Note that only a subset of the user's contacts is presented for each round to increase usability. The people's names are shown on mouseover. In order to increase the challenge, the composition of presented profile pictures depends on the current item's visibility settings. As an example, contacts and random strangers are equally distributed if the item is visible to all contacts. In contrast, if the item is only set to be visible to a specific set of contacts, then the composition of presented people largely uses the user's contacts.}. 

The subsequently described flow of Find Your Friends follows the experimental gaming model presented in the background section. The main \textit{challenge} for the learner is to demonstrate his privacy awareness, i.e. his knowledge of who can see his shared items. Following the experimental gaming model, the \textit{clear goal} for each round is to select those profile pictures that can see the presented item in order to reach the maximum score. Find Your Friend provides immediate \textit{feedback} based on the user's answer. A correctly selected profile picture is framed in green. Incorrect answers result in a red-framed profile picture. Additionally, the score is reduced by 1000 points and one of five hearts. A round is lost if either no heart is left or the score has fallen to zero points. Starting with a score of 10000 points for each round, the score is automatically reduced by 200 points each second to achieve \textit{focused attention}. With each round, the player gains new insights about his visibility settings, enabling him to gain \textit{control over the game} and reach higher scores.

It is notable that with increased privacy awareness, the game story shifts from a self-centred challenge to a community-centred one. Self-centred challenge refers to a player simply trying to learn who can see his shared items. With increasing control over the game, beating the score shared on Facebook by contacts becomes the main challenge instead of plain knowledge of the items' visibility. Yet, both challenges can be seen as motivational factors that contribute to the learning objectives.

\subsubsection{Score \& Feedback step}
After five rounds of Find Your Friends, Friend Inspector continues with the fourth step of the process flow (cf. Figure \ref{process}). The objectives of this step are to summarise the results of Find Your Friends, to calculate the overall score, and to provide personalised recommendations to improve privacy settings.

\begin{figure}[!h]
\centering
\includegraphics[width=3.40in]{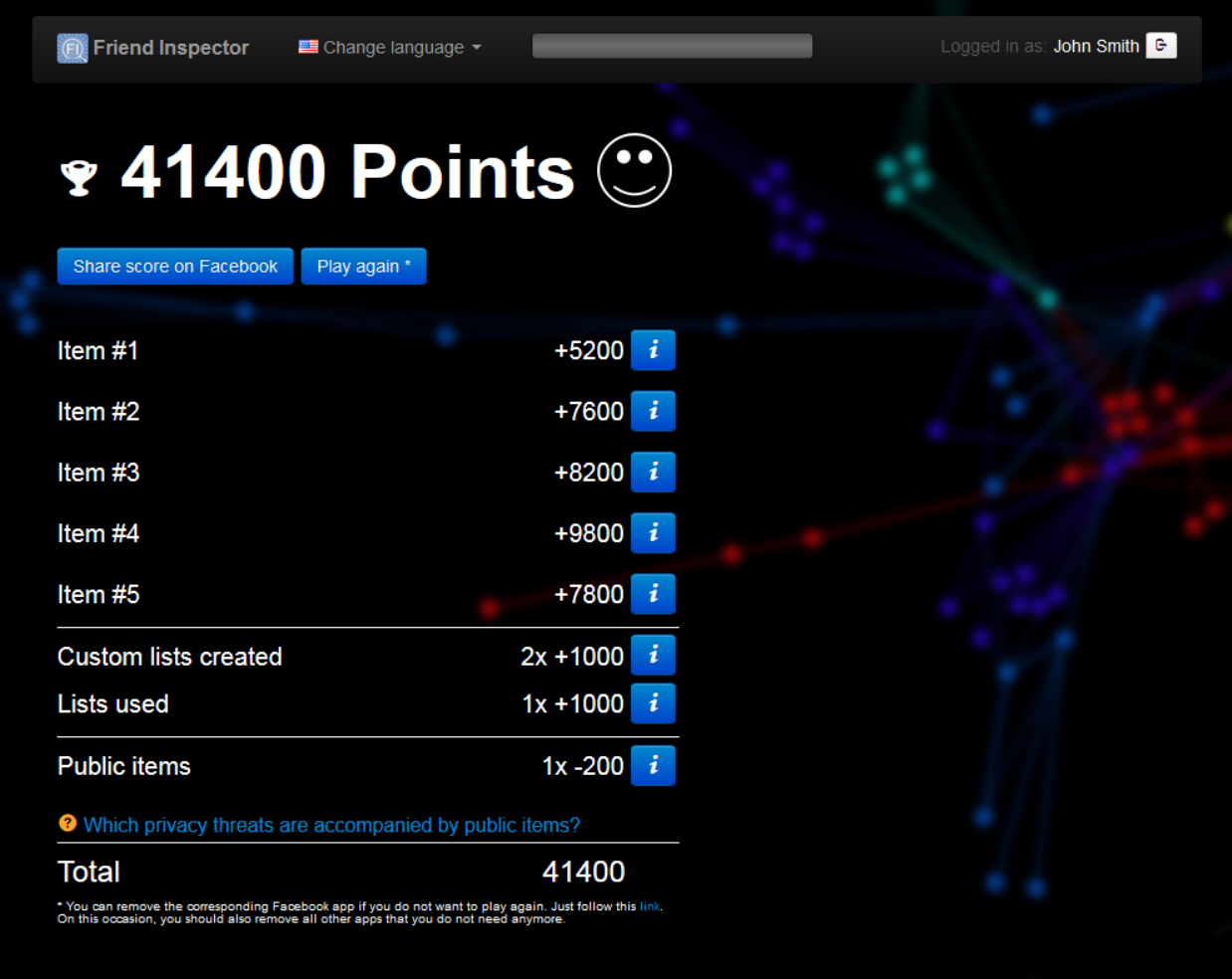}
\caption{Score \& Feedback (step 4)}
\label{step4}
\end{figure} 

Figure \ref{step4} depicts the Score \& Feedback interface. The upper part shows the overall score together with a smiley that allows the learner to easily put the score in context\footnote{Friend Inspector includes three different smileys: "Sad" smiley (\textless 15000 points), "Neutral" smiley (15000 -- 32500 points), "Happy" smiley (\textgreater 32500 points).}. Below the overall score, a detailed calculation allows the learner to understand how the score is composed of single item scores. Further information is available for each item through an expendable panel, showing incorrect answers. 

Additional bonus points are assigned for the definition and the use of friend lists on Facebook\footnote{Friend lists improve targeted sharing of items and thus contribute to privacy. For up to five lists, a user gets 1000 points for the definition and use of every list.}. Moreover, the player's score is reduced by 200 points for every item that is publicly shared. Reducing the score for publicly shared items raises the player's awareness for the high privacy risks of such a visibility setting.

Finally, based on the results and the user's Facebook profile, a set of personalised recommendations is displayed. Recommendations comprise instructions such as how to create friend lists, how to share personal items in a targeted manner, and how the term friendship on SNSs differs from friendships in the physical world. Following Friend Inspector's inductive learning, after experimenting with their privacy settings, these recommendations provide guidance and empower users to actually improve their privacy settings.

\section{Implementation and Dissemination}

Based on the conceptual design, we implemented Friend Inspector as a browser-based application which is publicly available\protect{\footnote{http://www.friend-inspector.org/}}. In order to offer an acceptable gaming experience, Friend Inspector imposes minimum requirements on the user's Facebook profile such as at least seven (non-public) shared items (pictures and status messages).

\subsection{Software architecture}

\begin{figure} [!t]
\centering
\includegraphics[width=1.0\columnwidth]{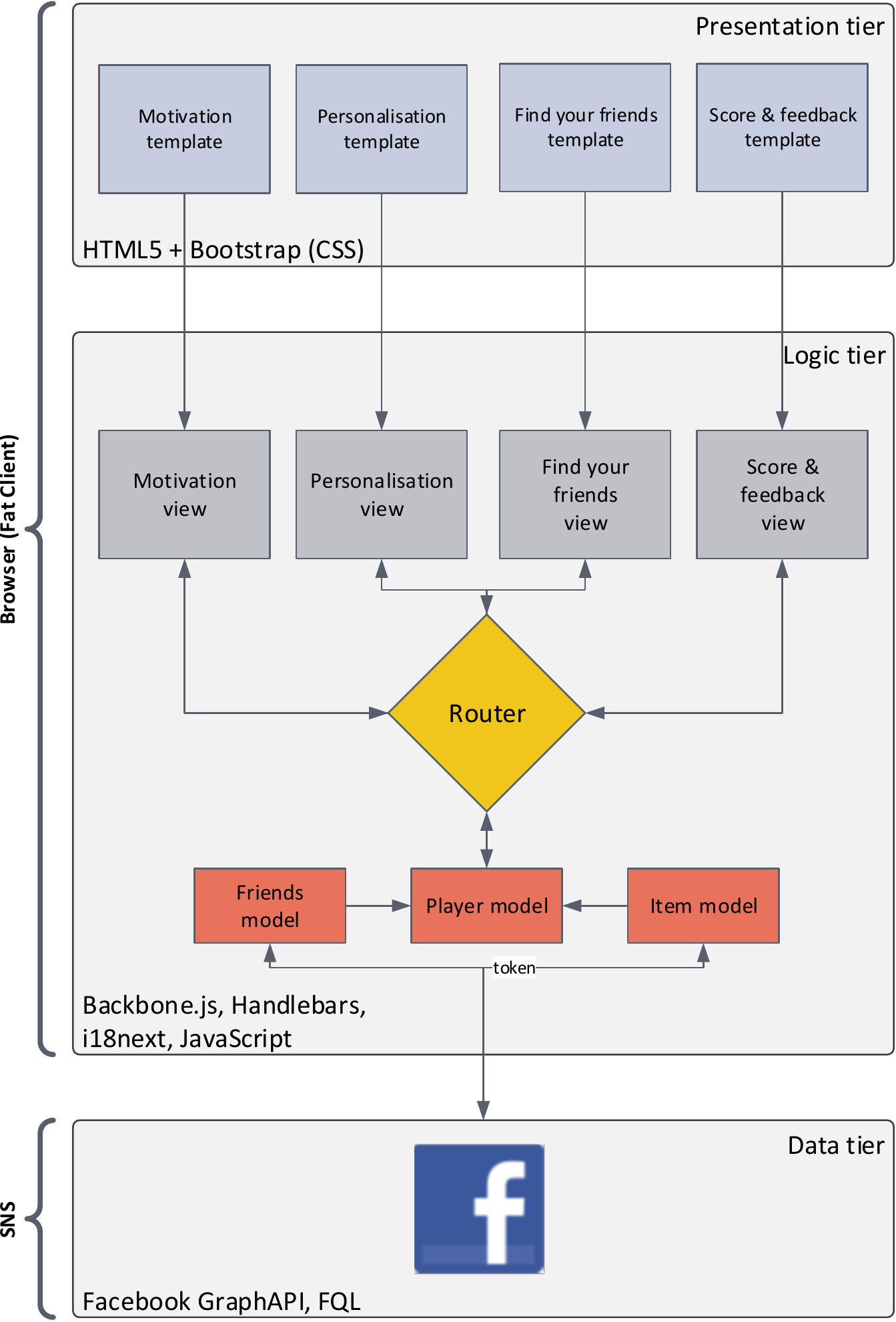}
\caption{Schematic view of Friend Inspector's architecture}
\label{fig:architecture}
\end{figure}

With respect to privacy issues regarding the software itself, Friend Inspector was developed as a client-side single-page application following a three-tier architecture. Thereby, the logic tier is designed according to a model-view-controler (MVC) approach. Figure \ref{fig:architecture} gives an overview of the architecture.

The data tier provides personal data, namely items, friends and privacy settings of a learner's profile. Since these private and sensitive data should not be stored persistently somewhere other than facebook, the GraphAPI serves as single persistent read-only data source/store. A connection to Facebook is established using the official Facebook SDK for JavaScript\footnote{https://developers.facebook.com/docs/reference/javascript/} (GraphAPI and FQL). In order to collect user data, the application must request a unique access token from Facebook, which is bound to the user, the application and the specific permission set.

The logic tier is completely implemented in JavaScript using backbone.js to enable the implementation of the whole logic functionality on the client-side (fat client). Backbone.js\footnote{http://backbonejs.org/} is a commonly used MVC framework that is suitable for the development of single-page websites. The framework involves models, collections, views and routers. The models (red rectangles in Figure \ref{fig:architecture}) represent Facebook entities as local transient instances. The router (orange rhombus) serves as a handler that guarantees a flawless process flow triggered by events. Moreover, it acts as a link between the models and the single views for each process step. The views (grey rectangles) render the allocated data and fill the templates (blue rectangles) of the presentation tier. For a clear separation of logic and design, we integrated the template engine Handlebars\footnote{http://handlebarsjs.com/}. 

The website presented to the user (presentation tier) is built on top of Bootstrap 2\footnote{http://getbootstrap.com/} and HTML5. Bootstrap is a widely used and recognized CSS framework that can improve the usability and trustworthiness of the game using a familiar layout theme.

\subsection{Secure and trusted implementation}
As Friend Inspector operates on the user's personal Facebook data, a secure and trusted implementation is of major importance to gain acceptance. Therefore, Friend Inspector is conceived as a client-side application that solely runs in the user's browser with no server-sided functionality required. As a result, personal data requested from the Facebook profile does not leave the user's domain at any time and is not transmitted to any server. Prior to the first use of Friend Inspector, users are informed about the game's access to their profile and must approve this action. This approach ensures ethical data capture and use of personal information within the game. Additionally, the source code of Friend Inspector has been released under the Apache License 2.0 and is publicly available to further increase trust in the application, allowing interested users to verify its integrity and security.

\subsection{Dissemination and usage}
Friend Inspector is aimed to raise privacy awareness of as many SNS users as possible. Consequently, disseminating the game has been of major importance. To this end, Friend Inspector uses i18next\footnote{http://www.i18next.com/} to add multilanguage support, which also allows to add easily new language files. Friend Inspector is currently available in two languages (English and German). 

Friend Inspector was launched on June 26\up{th}, 2013 and has been widely covered in national and international media (press, radio, and television). Within five months, the Friend Inspector site has been requested more than 100,000 times. Note that detailed usage statistics are not available, as Friend Inspector does neither store nor analyse log files for privacy reasons. Amazon's Elastic Compute Cloud\protect{\footnote{http://aws.amazon.com/en/ec2/}} (Ireland) is used to host Friend Inspector for performance reasons and to cope with traffic peaks.

\section{conclusions}

In this paper, we introduced Friend Inspector, a serious game developed to enhance SNS users' privacy awareness. Friend Inspector addresses the current challenge of SNS users, namely to understand who can see their shared personal items. In order to address especially younger users and protect them from social exclusion due to the consequences of poor privacy awareness, a game-based approach has been chosen. The conceptual design of Friend Inspector is based on two foundations: firstly, an in-depth understanding of privacy awareness as the match or mismatch between perceived and actual visibility of shared items. Secondly, an inductive learning approach that allows its users to experiment and play with their own Facebook data in order to actively learn about the visibility of their personal items.




Friend Inspector is implemented as a web application for the SNS Facebook. In the five months since its launch, Friend Inspector has attracted a significant number of visitors, which further emphasises the need for better tools to understand privacy settings on SNSs.

\section{Acknowledgements}
Research conducted by Moritz Riesner was supported by "Regionale Wettbewerbsf\"{a}higkeit und Besch\"{a}ftigung", Bayern, 2007-2013 (EFRE) as part of the SECBIT project (http://www.secbit.de/). Research conducted by Michael Netter and Johannes S\"{a}nger was supported by "Bavarian State Ministry of Education, Science and the Arts" as part of the FORSEC research association (http://www.bayforsec.de/).

\balance

\bibliographystyle{acm-sigchi}
\bibliography{references}
\end{document}